\documentclass[iop]{emulateapj}
\shorttitle{Searching for Binary Y dwarfs with GeMS}
\shortauthors{Opitz et al.}
\begin{document}
\title{Searching for Binary Y dwarfs with the Gemini Multi-Conjugate Adaptive Optics System (GeMS)}
\author{Daniela Opitz\altaffilmark{a,b}, C. G. Tinney\altaffilmark{a,b}, Jacqueline K. Faherty\altaffilmark{c,d,e}, Sarah Sweet\altaffilmark{f}, Christopher R. Gelino\altaffilmark{g,h}, J. Davy Kirkpatrick\altaffilmark{g}}
\email{daniela.opitz@student.unsw.edu.au}
\altaffiltext{a}{School of Physics, University of New South Wales, NSW 2052, Australia}
\altaffiltext{b}{Australian Centre for Astrobiology, University of New South Wales, NSW 2052, Australia}
\altaffiltext{c}{Department of Terrestrial Magnetism, Carnegie Institution of Washington, Washington, DC 20015, USA}
\altaffiltext{d}{Department of Astrophysics, American Museum of Natural History, Central Park West at 79th Street, New York, NY 10034, USA}
\altaffiltext{e}{Hubble Fellow}
\altaffiltext{f}{Research School of Astronomy \& Astrophysics, Australian National University, Canberra, ACT 2611, Australia}
\altaffiltext{g}{Infrared Processing and Analysis Center, MS 100-22, California Institute of Technology, Pasadena, CA 91125, USA}
\altaffiltext{h}{NASA Exoplanet Science Institute, MS 100-22, California Institute of Technology, Pasadena, CA 91125, USA}

\begin{abstract}
The NASA Wide-field Infrared Survey Explorer {\em (WISE)} has discovered almost all the known members of the new class of Y-type brown dwarfs. Most of these Y dwarfs have been identified as isolated objects in the field. It is known that binaries with L- and T-type brown dwarf primaries are less prevalent than either M-dwarf or solar-type primaries, they tend to have smaller separations and are more frequently detected in near-equal mass configurations.  The binary statistics for Y-type brown dwarfs, however, are sparse, and so it is unclear if the same trends that hold for L- and T-type brown dwarfs also hold for Y-type ones. In addition, the detection of binary companions to very cool Y dwarfs may well be the best means available for discovering even colder objects. We present results for  binary properties of a sample of five {\em WISE} Y dwarfs with the Gemini Multi-Conjugate Adaptive Optics System (GeMS). We find no evidence for binary companions in these data, which suggests these systems are not equal-luminosity (or equal-mass) binaries with separations larger than $\sim$ 0.5-1.9 AU. For equal-mass binaries at an age of 5 Gyr, we find that the binary binding energies ruled out by our observations (i.e. $10^{42}$\,erg) are consistent with those observed in previous studies of hotter ultra-cool dwarfs. 
\end{abstract}

\section{Introduction}

The discovery of the coolest Y spectral class of brown dwarfs has extended the temperature range for isolated star-like objects down to \textbf{$\sim$} 250 K \index{Y dwarfs} \citep{CUSHING11, LUHMAN14}.
Their discovery enables the study of the properties of objects in the temperature gap between the coolest previously known sub-stellar objects (T$_{eff}$ $\sim$ 500\,K) and gas-giant planets (T$_{eff}$ $\sim$ 130\,K). Currently twenty--one Y dwarfs are known \citep{CUSHING11, KIRK12, TINNEY12, LIU12, KIRK13, CUSHING14, PINFIELD14}, as well as three candidates awaiting spectroscopic confirmation \citep{LIU11, LUHMAN11,LUHMAN14,SCHNEIDER15}.

Most of the spectroscopically confirmed Y dwarfs  have been  identified as isolated field objects by the NASA Wide-field Infrared Survey Explorer \citep[\textit{WISE},][]{WRIGHT10}. Searches for very low-mass binaries (defined as having a total system mass M$_{tot}$ $<$ 0.2 M$_\odot$ and primary mass M$_1$ $<$ 0.1 M$_{\odot}$) have concentrated on high resolution imaging surveys, using both nearby field sources \citep[e.g; ][]{KOERNER99,BURGASSER03,REID08,GELINO2011,ABERA14} and young cluster associations \citep[e.g;][]{MARTIN98,NEUHAUSER02,BOUY06,TODOROV14}. These studies have determined a brown dwarf binary fraction of $\sim$ 10\%--30\% \citep{BURGASSER07}, which is substantially lower than the binary fraction of solar-type stellar systems \citep[$\sim$ 65\%; ][]{DUQUENNOY91} and the binary fraction of early-type M stars \citep[$\sim$ 30\%--40\%;][]{DELFOSSE04,REID97}. This trend could indicate either a mass dependence on the multiplicity or an as yet uncovered population of very low-mass binaries. The latter is strongly supported by the known incompleteness of the statistics for very tight (\textit{a} $\lesssim$ 1 AU) and wide (\textit{a} $\gtrsim$ 100 AU) binaries \citep[see][and references therein]{KONOPACKY13}.

The binary status of Y type brown dwarfs is also both unclear and of considerable interest. Open questions include: Is there a lower mass limit for the formation of binary systems? How common are Y dwarf binary systems? What is the mass ratio distribution between the components of Y dwarf binaries? A new generation of wide-field adaptive optics systems using laser-guide star constellations and deformable mirrors conjugating to multiple layers in the atmosphere offer the prospect of addressing these questions from the ground (in advance of JWST's capabilities becoming available in space).

Binarity, in addition, has been proposed as an explanation for some of the spread seen in the absolute magnitudes of otherwise similar Y dwarfs \citep{TINNEY14, LEGGETT15}. The latest atmospheric models \citep{MORLEY12} are consistent with the majority of the observed absolute magnitudes for Y dwarfs. However, some (including WISEA\,J053516.87--750024.6 and WISEA\,J035934.07--540154.8 studied in this paper) show disparities. These objects appear to be over-luminous in MJ and MW2 relative to cloud-free models suggesting either the presence of condensate clouds or equal-mass binarity. Binary Y dwarf systems, once identified, also offer the opportunity to empirically measure dynamical masses \citep[e.g;][]{DUPUY09, KONOPACKY10}.

These issues motivated a diffraction-limited study to determine the binary status of five Y dwarfs using the Gemini  Multi-Conjugate Adaptive Optics System (GeMS). 
In Section 2, the properties of our sample, observations and data reduction are detailed. In Section 3 the binary status of our targets is examined, and conclusions are presented in Section 4.

\section{Observations and data reduction}

\begin{deluxetable*}{cccccccc}
\tablecolumns{4}
\tablecaption{Y dwarf sample \label{tab:table1}}
\tablehead{
\colhead{Full designation}  & \colhead{Short name}  & \colhead{J3}     & \colhead{$\pi$}   & \colhead{Spectral type}\\ 
                            &                  & \colhead{(mag)}  & \colhead{(mas)}   &      & }
\startdata
WISEA J035934.07--540154.8 & W0359 & 21.40 $\pm$ 0.09 & 63.2 $\pm$ 6.0  & Y0        \\ 
WISEA J053516.87--750024.6 & W0535 & 22.09 $\pm$ 0.07 & 74.0 $\pm$ 14.0 & $\geq$Y1 \\ 
WISEA J071322.55--291752.0 & W0713 & 19.42 $\pm$ 0.03 & 108.7 $\pm$ 4.0 & Y0        \\ 
WISEP J154151.65--225025.2 & W1541 & 20.99 $\pm$ 0.03 & 175.1 $\pm$ 4.4 & Y1        \\
WISEA J163940.83--684738.6 & W1639 & 20.57 $\pm$ 0.05 & 202.3 $\pm$ 3.1 & Y0pec     \\ 
\enddata

\tablecomments{Target magnitudes are provided in the J3 passband (1.29 $\mu$m) as described in \cite{TINNEY14}, along with parallaxes from the same source. Spectral types are from: \citet{KIRK13} and \cite{SCHNEIDER15}.}
\end{deluxetable*}

We observed a sample of five nearby Y dwarfs discovered by {\em WISE} (see Table \ref{tab:table1}).  The full {\em WISE} designation, near-infrared J3 photometry, parallax and spectral type for the  sample are also listed. These objects were observed with the Gemini South Adaptive Optics Imager \cite[GSAOI,][]{MCGREGOR04, CARRASCO12} and corrected for atmospheric aberrations by the Gemini Multi--Conjugate Adaptive Optics System \cite[GeMS,][]{ORGEVILLE12}. GSAOI has a pixel scale of 0.02$\arcsec$ and is composed of four 2048$\times$2048 Rockwell HAWAII--2RG  arrays that form a near-infared imaging mosaic. Each  detector offers access to a field of view of 41$\arcsec$x41$\arcsec$. All observations were carried out in the GSAOI $CH_{4}S$ passband (1.486--1.628\,$\mu$m). This filter was chosen as this provides the optimal sensitivity for these faint objects with very strong methane absorption.

The extreme faintness of Y brown dwarfs combined with the rarity of suitably bright natural guide stars makes natural guide star adaptive optics for these targets completely impractical. The GeMS system was chosen for these observations over a traditional single-deformable mirror system, because its  wide field of correction allows the selection of off-axis tip-tilt stars over a large field, as well as delivering AO correction over a large $\approx$ 2' diameter field. This to-date unique capability allows observations of Y dwarfs to address both ``narrow field'' binarity science, as presented here, and wide-field astrometric science, to be presented in a future publication.

A log listing the observations is given in Table~\ref{tab:table2}. The Y dwarfs W1541, W0713 and W1639 were observed between March 2013 and May 2013 with a total integration time of approximately 1 hour, using 54 exposures of 66s each and random telescope dithering every 6 exposures inside a box size of $\sim$1.6\arcsec$\times$1.6\arcsec.

Experience with this observing mode showed that observing overheads were high and therefore, subsequent observations for W0359 and W0535 were carried out by dithering and co-adding every 9 exposures. These observations delivered a typical FWHM of 86 mas for W1541, W0713 and W1639 and a FWHM of 120 mas for W0359 and W0535. The difference in the FWHM for these two groups of objects are caused by the different observing conditions.

Data processing was performed using the Gemini GSAOI pipeline, which operates in the {\em IRAF} environment\footnote{http://www.gemini.edu/sciops/data--and--results/getting--started\#gsaoi}. 
This applies a bad pixel mask, creates and subtracts an averaged dark from all images, applies a flat-field generated using dome flats and generates sky frames using dithered data sets which are then subtracted. Finally dithered images are combined using reference stars to produce a single mosaicked image. 

\begin{figure*}
\centering
\hspace*{-1.6in}
\includegraphics[angle=0, scale=1.0]{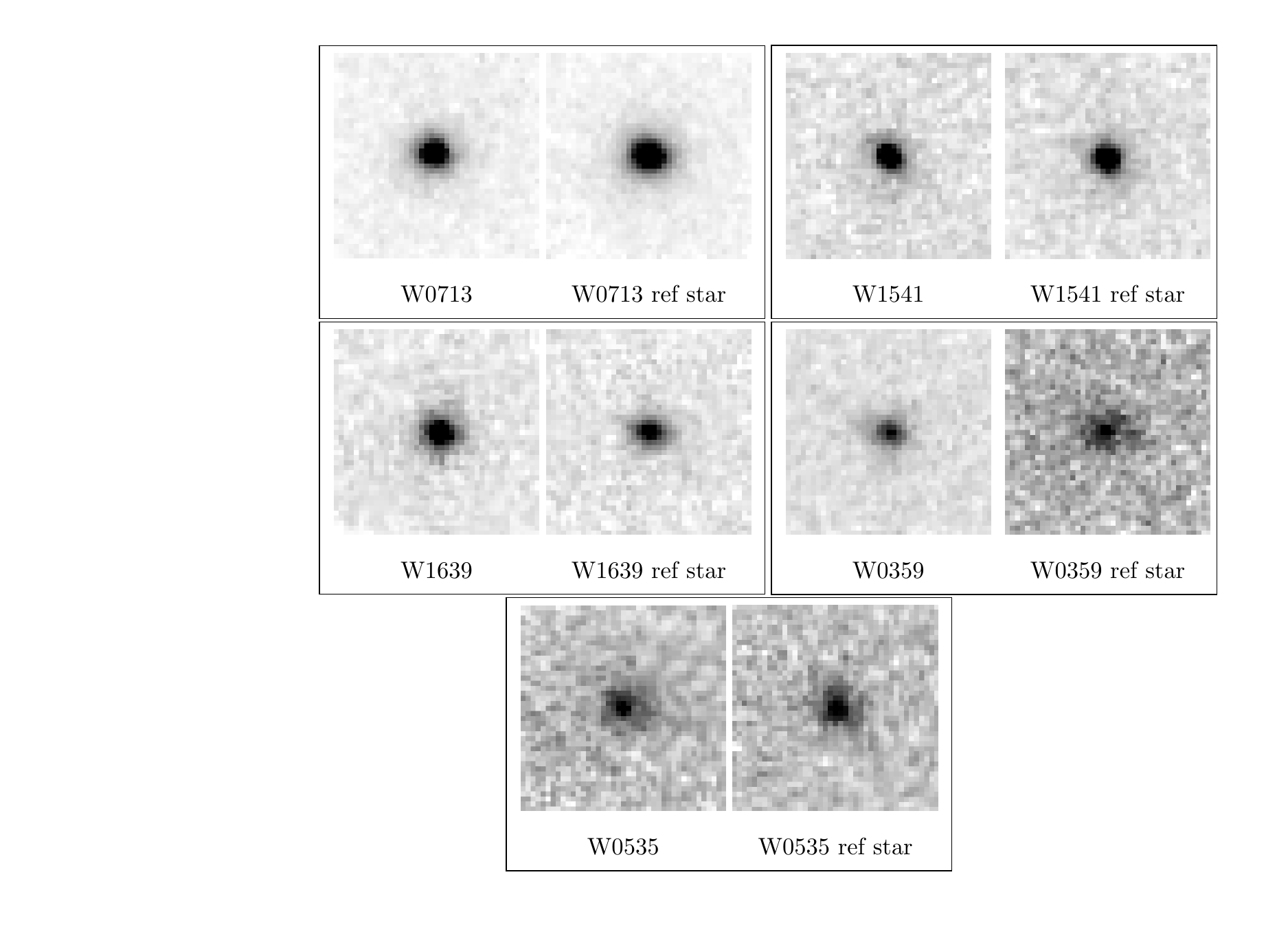}
\caption{\textit{Left}:  Y dwarfs imaged in the  CH$_4$s filter. For each object two panels are shown, with the left displaying a 0.8\arcsec$\times$0.8\arcsec\ sub-image centered on the Y dwarf, and the right showing a nearby reference star. All images are north up and east to the left.}
\label{fig:Images}
\end{figure*}

The creation of the final mosaicked image relies on the presence of sufficient reference stars in the field to perform an astrometric registration. For four of our targets (W0535, W0713, W1541, W1639) this analysis could be done for the single detector containing our target. However, for W0359 we needed to process all four detectors together to make a 2$\times$2 mosaic in order to acquire sufficient reference stars for this step. The FWHM in the final mosaics for each target, determined from Point Spread Function (PSF) analysis described in Section 3,  are listed in Table~\ref{tab:table2}.
Postage-stamp images zooming on a 0.8"$\times$0.8" region around each of our targets along with a nearby unresolved reference star are shown in Figure~\ref{fig:Images}. These show that no obvious binary companions are found in these data.

\begin{deluxetable*}{ccccccc}
\tablecolumns{7}
\tablecaption{Log of GSAOI--GeMS Observations.\label{tab:table2}}
\tablehead{
\colhead{Short} & \colhead{UT Date}   & \colhead{Exp.} & \colhead{Array$^1$} & \colhead{Gain}          & \colhead{FWHM} &\colhead{FWHM}\\ 
\colhead{name} &                               & \colhead{(s)}   &                         & \colhead{ (e-/ADU)}  & \colhead{(pix)} & \colhead{($\arcsec$)} }
\startdata
W0359          & 2013 Dec 20  &  360s$\times$9          & 3   & 2.41  & 5.51  & 0.11 \\ 
               &              &                         &     &       &       &      \\ 
W0535          & 2014 Dec 04  &  360s$\times$9          & 3   & 2.41  & 6.37  & 0.12 \\ 
               &              &                         &     &       &       &      \\ 
\vspace{-0.2cm}& 2013 Mar 22  & 66s$\times$54$\times$9                        &     &       & 3.99  & 0.08 \\
W0713          &              &                         & 2   & 2.01  &       &      \\
               & 2013 Apr 20  & 66s$\times$54$\times$9                       &     &       & 4.47  & 0.09 \\  
               &              &                         &     &       &       &      \\ 
\vspace{-0.2cm}& 2013 Apr 20  & 66s$\times$54$\times$9  &     &       & 4.13  & 0.08     \\
W1541          &              &                         & 2   & 2.01  &       &          \\
               & 2013 May 24  &  66s$\times$54$\times$9 &     &       & 3.98  & 0.08      \\
               &              &                         &     &       &       &           \\ 
W1639          & 2013 Apr 21  &  66s$\times$54$\times$9 & 4   & 2.64  & 4.92  & 0.09      \\
\enddata
\tablecomments{1 --Mosaic detector where each target is located. For W0359 we processed all the GSAOI arrays into a combined mosaic. For the  other targets only the individual detector with the target was processed to a final image. }
\end{deluxetable*}

\section{Binary Analysis}

Figure~\ref{fig:Images} shows no obvious evidence for close binary companions in our data. To examine this more closely we have performed two analyses to understand the presence (or absence) of binary companions. We concentrate in this study on close binary companions –- the presence (or absence) of wider companions is better probed by multiple epochs of natural-seeing data (since the confirmation of wider companions will critically rely on the observations of common proper motions) and is therefore deferred to a future paper.

\subsection{Point Spread Function Analysis}
\label{subsec:psf}

We obtained a Point Spread Function (PSF) model for each image using the DAOPHOT II package \citep{STETSON87} implemented within the Starlink\footnote{http://www.star.bris.ac.uk/$\sim$mbt/daophot/} environment.
Unsaturated stars were selected and used to determine an initial model PSF, which was used to fit and subtract all known stars within each image in a first pass processing. Any objects detected in the first-pass PSF-subtracted image (and in particular any objects detected near the PSF stars) are added into the list of known stars and included in a second pass of analysis, so as to iterate toward an uncontaminated single-star PSF. This final PSF  was then used to simultaneously fit to all objects in the field, allowing a final subtraction of all known objects from the mosaics.
This process did not reveal any companions within $\approx$1.0\arcsec\ of our targets. It also yielded a PSF model for each image that was used in subsequent image injection simulations.

\subsection{Companion Detection Simulations}
\label{subsec:limits}

We used to two methods to explore the detectability of potential binary companions of our target stars, and to determine the magnitude-difference and separation-limits implied by our non-discovery of companions.

\subsubsection{Artificial Star Injection Simulations}

This method injects synthetic binaries with a variety of separations and magnitude differences into the images, then treats these new systems as both single and binary systems and fits PSF models to them.

We first construct a 160x160 pixels sub-image centered on each Y dwarf. Into those sub-images, we inject a pair of synthetic stars at 4 positions 1.13'' away from the Y dwarf at position angles of 45\arcdeg, 135\arcdeg, 225\arcdeg\ and 315\arcdeg\ (i.e. pixel positions (40,40), (40,120), (120,40) \& (120,120) in the sub-image). This radial separation from the Y dwarf is small enough that the injected systems have the same sky background as the actual Y dwarf, and far enough away that they are uncontaminated by the Y dwarf. (The exception to this is the W1541 data which has a bright star that contaminated the (120,120) position, so it was moved to an offset of (130,100)). The synthetic binaries were injected with radial separations of 1,2,3, .. 10 pixels at positions angles of 0--360\arcdeg\ in steps of 45\arcdeg, and with magnitude differences ($\bigtriangleup$mag) corresponding to flux ratios of 1.00--0.05. In total we injected 2280 synthetic binary systems into each Y dwarf image.

After the injection of artificial binaries, we used DAOPHOT to fit both single and binary models \cite[generally following the analysis used by ][]{ABERA14}. We made an initial guess for the position of the primary (by detecting a peak identified in the region of the injected stars and fitting to it as a single star) and the secondary (by detecting a peak in the residual image obtained after subtracting the first object detected), and then used DAOPHOT to fit for both a single and a binary model. An illustration of this process is shown in Fig.~\ref{fig:Image2}.

The relative statistical significance of the single-star and binary-star fits was assessed using the one-sided F-test

\begin{equation}
\label{eq:ftest}
F={{\chi_{\mathrm{sin}}/\nu_{\mathrm{sin}} }\over{\chi_{\mathrm{bin}}}/\nu_{\mathrm{bin}}}
\end{equation}

where $\chi_{\mathrm{sin}}$ and $\chi_{\mathrm{bin}}$ are the usually defined $\surd\chi^2$ for each model fit, and $\nu_{\mathrm{sin}}$ and $\nu_{\mathrm{bin}}$ are the degrees of freedom for a single and binary model fit. The latter  were computed using the following expressions:

 \begin{equation}
\label{eq:ftest2}
{\nu_{sin/bin}= pix_{\mathrm{eff}}} - N
\end{equation}
 
where $pix_{\mathrm{eff}}$ are the ``effective pixels'' involved in each fit (essentially a normalised measure of the number of pixels meaningfully involved in each fit - see \cite{ABERA14}) and N is the number of parameters for the model (3 for a single star and 6 for a binary). A significance level ($\alpha$) of 0.05 was required to pass this test -- i.e we are required to have more than 95\% confidence that the binary model is preferred over the single star model. We then use the 2280 synthetic binary systems to determine the separations and magnitude-differences at which $>$50\% and $>$90\% of injected binaries were recovered with 95\% confidence. These results are plotted in Fig. \ref{fig:limit1} for each independent observation, with grey symbols showing the 50\% confidence curves and blue symbols showing the 90\% confidence curves. A counter-intuitive feature of these curves \cite[also seen by][]{ABERA14} is that equal-luminosity binaries are {\em slightly} harder to detect than slightly non-equal luminosity binaries.

\begin{deluxetable}{ccc}
\tablecolumns{3}
\tablecaption{Signal-to-Noise Data\label{tab:SN}}
\tablehead{
\colhead{Short name}  & \colhead{S/N}  & \colhead{S/N}      \\ 
                      &                & \colhead{(ref star)}  }
\startdata
W0713 & 200   & 166.67      \\ 
W1541 & 58.82 & 76.92       \\ 
W1639 & 76.92 & 52.63       \\ 
W0535 & 21.28 & 29.41       \\
W0359 & 37.04 & 18.52       \\ 
\enddata
\tablecomments{Estimate of the S/N ratios obtained for the Y dwarfs and reference stars displayed in Fig. ~\ref{fig:Images}.}\end{deluxetable}

\textbf{\begin{figure}
\centering
\includegraphics[angle=0, scale=1]{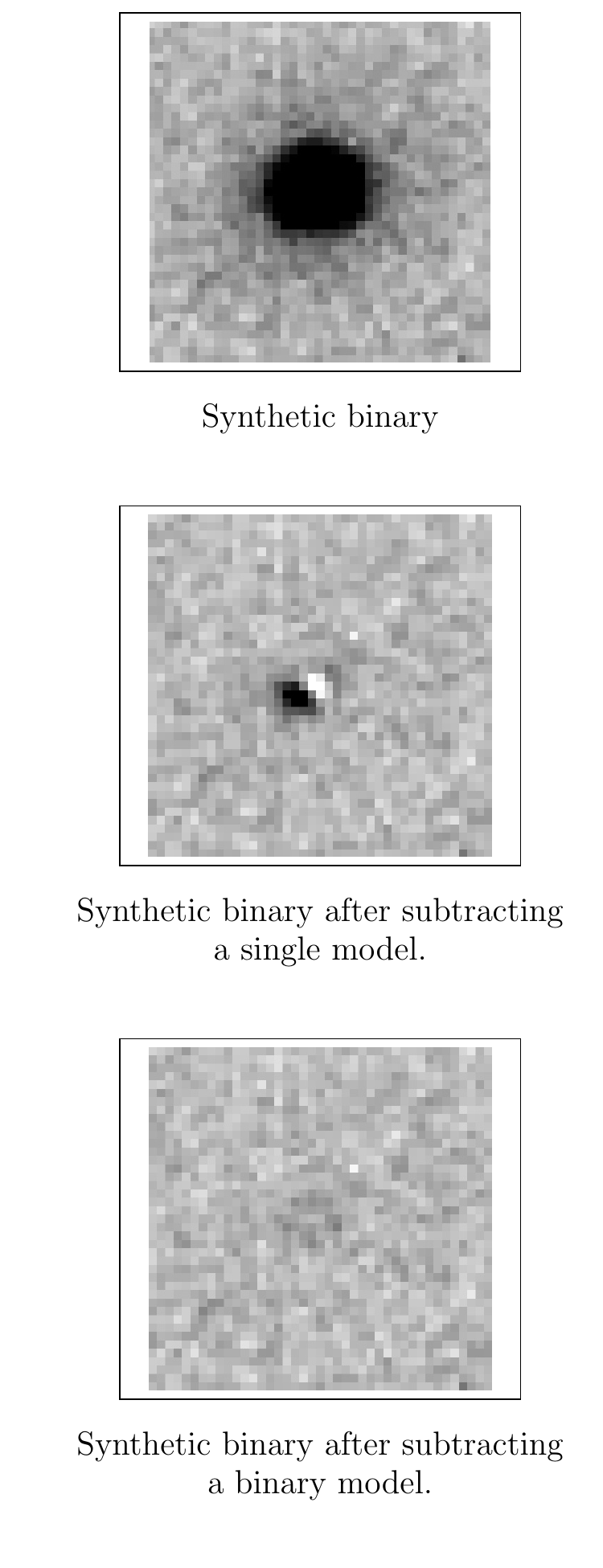}
\caption{Synthetic Y dwarf binary with a $\bigtriangleup$mag = 2.0 created using a W0713 point-spread function. The upper panel shows the binary at a separation of 3 pixels (0.06\arcsec). The middle panel displays the same binary after the subtraction of a single PSF model, displaying a clear residual source. The bottom panel displays the binary after the subtraction of a binary model. All images are at the same gray scale and show a 0.8\arcsec$\times$0.8\arcsec\ region on a side with north up and east to the left.}
\label{fig:Image2}
\end{figure}}

\subsubsection{Photon-Counting}

The artificial star injection simulations above show that (in general) binary companions up to 1.5 magnitudes fainter than the primary can be detected to within one half of an image FWHM. To explore the sensitivity of our data to wider (and fainter) companions, we compute the flux a hypothetical companion would need to have in order to be detected with S/N=3  (or equivalently photometric uncertainty 0.33 mag) at a series of annular radii from the brown dwarf \citep[following the analyses performed by ][]{GELINO2011}. To estimate these limits we constructed a set of 50 annuli with 1 pixel of width and radius between 3 \& 50 pixels from the brown dwarfs. We computed the standard deviation of the pixel values in each annulus ($\sigma_a$). Then, we estimated the flux F a companion would have from the standard equations for the magnitude uncertainty, 

\begin{equation}
\label{eq:sigma1}
\sigma_m={{C*\sigma_F }\over{F}}
\end{equation}

\begin{equation}
\label{eq:sigma2}
\sigma_F=\sqrt{(A*\sigma_a^2) + (A^2*\sigma_a^2/N_a) + F/G}
\end{equation}

\noindent where C is a constant equal to 1.0857, G is the gain of the detector (see Table~\ref{tab:table2}), $\sigma_m$ is 0.33, the magnitude error for detection limits at 3$\sigma$ over the sky, A is the area of the aperture for the detection of the companion ($\pi*[3$ pixels $]^2$) and N$_a$ is the number of pixels in the sky annulus.

Combining \ref{eq:sigma1} and \ref{eq:sigma2} and solving for F results in:

\begin{equation}
\label{eq:flux}
{\scriptstyle
F ={{C^2/(G*\sigma_m^2)}+\sqrt{(C^2/(G*\sigma_m^2))^2 + 4*(A*\sigma_{a}^2+ A^2*\sigma_a^2/N_{a})*C^2/\sigma_m^2}\over{2} }}
\end{equation}

The flux F of the hypothetical secondary was then converted to a magnitude ($m_2$) using the standard equation

\begin{equation}
\label{eq:mag}
m_2 \propto -2.5*log_{10}(F)
\end{equation}

and the magnitude difference was computed as the difference between  $m_2$ and the magnitude of the target ($m_T$). 
DAOPHOT is occasionally unable to determine a reliable modal sky value. When this happens we simply discount that $\sigma_a$ and its trial radii. The resulting separation- and magnitude-difference-limits are shown in Fig. \ref{fig:limit1} as red symbols.

\subsection{Results}

\begin{figure*}
\centering
\includegraphics[scale=0.875]{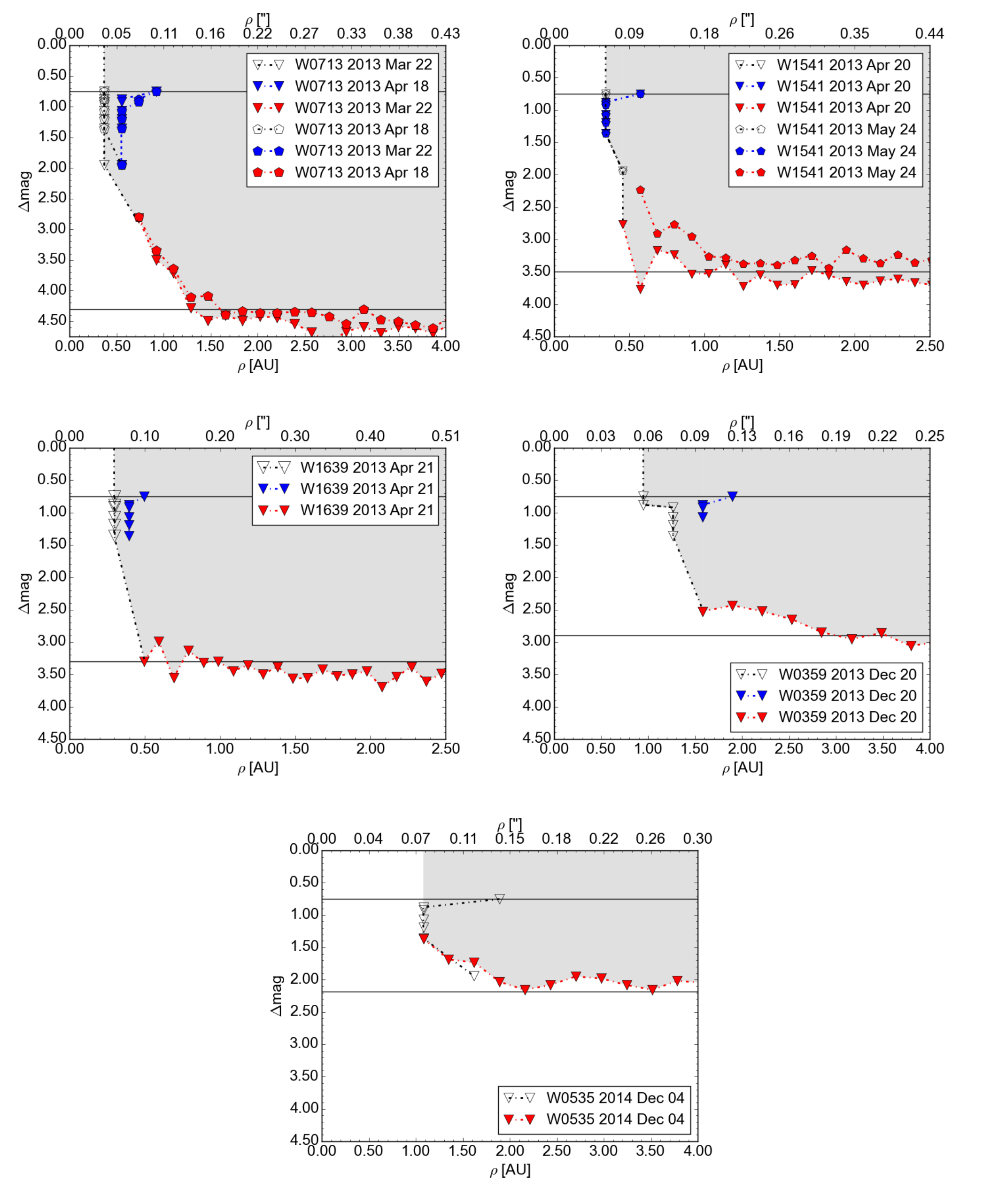}
\caption{Magnitude difference ($\bigtriangleup$mag) and separation ($\rho$) limits for five Y dwarfs at which 50\% (grey symbols) and 90\% (blue symbols) of companions were recovered, as computed using artificial star injection. The horizontal lines intersect the magnitude axis at 0.75 mag (corresponding to an equal-luminosity binary) and at the maximum $\bigtriangleup$m.}
\label{fig:limit1}
\end{figure*}

The regions of magnitude-difference versus separation space ruled out by these observations are shown in Fig.~\ref{fig:limit1}, with Table~\ref{tab:results} summarizing some key features of these diagrams -- namely the largest separation allowed for an unresolved equal-luminosity binary, and the largest magnitude difference ruled out by these data.

As a general rule, the artificial star injection technique is more powerful at small radial separations, where the Y dwarf is imaged with good S/N. In this case an accurate model of the PSF is critical for determining the ability to resolve two closely separated targets. The photon-counting technique readily extends to large separations, and so estimates the faintest companion that our data can rule out. Our artificial injection simulations with a recovery fraction of 90\% allow us to strongly conclude that none of these Y dwarfs are equal-mass/equal-luminosity binaries with separations larger than $\sim$ 0.5-1.9 AU. These limits can be slightly extended for a less-confident recovery fraction of 50\% to $\sim$ 0.3-1.9 AU. Our best data is for W0713 and it shows no evidence for binarity to limits $\sim$ $\bigtriangleup$mag = 4.4 mag  at separations beyond 1.7 AU (0.18$\arcsec$).

\newpage
\begin{deluxetable*}{cccccc|cc|cc|c}[!ht]
\tablecolumns{10}
\tablecaption{Limits in separation and magnitude \\ from PSF injecting and photon--counting techniques. \label{tab:results}}
\tablehead{
\colhead{Short} & \colhead{UT Date} & \colhead{N$_{\mathrm{PSF}}$\tablenotemark{1}} & \colhead{$\bigtriangleup$mag\tablenotemark{2}} 
& \multicolumn{2}{c}{$\rho_{\mathrm{phot}}$\tablenotemark{3}} 
& \multicolumn{2}{c}{$\rho_{\mathrm{inj,50}}$\tablenotemark{4}} 
& \multicolumn{2}{c}{$\rho_{\mathrm{inj,90}}$\tablenotemark{5}}\\ 
\colhead{name}&                     & \colhead{}  &  & \colhead{($\arcsec$)} &  \colhead{(AU)} & ($\arcsec$) &\colhead{(AU)} &($\arcsec$)& \colhead{(AU)} }
\startdata

      & 2013 Mar 22 &      			& 4.28    & 0.14       & 1.29                           & 0.04 & 0.37 &0.10 & 0.92&\\ 
W0713 &             & 13              &            &               &                                &      &     &&&\\ 
      & 2013 Apr 18 &      			& 4.39    & 0.18       & 1.66                           & 0.04 & 0.37&0.10&0.92&\\ 
      &             &                &       &            &                                &      &    &&& \\ 
      & 2013 Apr 20 &      			 & 3.53    & 0.16       & 0.91                           & 0.06 & 0.34&0.10&0.57&\\
W1541 &             & 10               &            &               &                                &      &   &&&  \\
      & 2013 May 24 &          	     & 3.28    & 0.18       & 1.03                           & 0.06 & 0.34&0.08&0.48&\\
      &             &                &        &            &                                &      &  &&&   \\ 
      &             &                &        &            &                                &      &   &&&  \\
W1639 & 2013 Apr 21 & 18   	  & 3.30   & 0.08       & 0.47                           & 0.06 & 0.30&0.10&0.49&\\
      &             &                &        &            &                                &      &  &&&   \\
W0359 & 2013 Dec 20 &  6    	 & 2.84   & 0.18        & 2.84                         & 0.06 & 0.95&0.12&1.90&\\ 
      &             &                &        &            &                                &      &   &&&  \\ 
      &             &                &        &            &                                &      &    &&& \\
W0535 & 2014 Dec 20 & 19   	& 2.16   & 0.16      & 2.16                           & 0.14 & 1.89&-&-&\\
\enddata
\tablecomments {1 -- Number of stars used to create the the point-spread function model. 2 -- Magnitude difference limits computed by the photon counting method. 3 -- Limits in separation between the primary and the secondary at the magnitude difference limit in arcsec and astronomical units respectively computed by the photon counting method. 4 -- Limits in separation for an equal-mass/equal-luminosity binary (magnitude difference of 0.75 mag) computed by the PSF injection method (for 50\% of objects recovered). 5 -- Limits in separation for an equal-mass/equal-luminosity binary (magnitude difference of 0.75 mag) computed by the PSF injection method (for 90\% of objects recovered).}
\end{deluxetable*}
\label{subsec:results}

\subsection{Equal mass/luminosity binaries}

\subsubsection{WISE 1541, WISE 0713 and WISE 1639}

WISE 0713, WISE 1541 and WISE 1639 are the objects for which our GeMS images have the highest quality, and the artificial star injection simulations show, with high levels of confidence (i.e. 90\%), that none of these are equal-mass/equal-luminosity binaries with separations larger than 0.5-1.9\,AU. Using a weaker confidence limit on binary recovery (50\%), we get only a {\em  slightly} tighter range of limits on separation.

\subsubsection{WISE 0535 and WISE 0359}

Recently \cite{TINNEY14} have highlighted a handful of Y dwarfs that are over-luminous in J$_{MKO}$ and W2 \cite[4.6 $\mu$m,][]{WRIGHT10} relative to cloud-free models. W0359 and W0535, in particular, were highlighted as being over-luminous by 0.6 and 1.1 mag respectively. If their over-luminosity were due to unresolved multiplicity, they would have to be nearly equal-luminosity/equal-mass binaries or triples. 

The Y--dwarfs W0359 and W0535 are the faintest in our sample, and so the most challenging targets for measuring binary limits. From the artificial injection simulations,  with a recovery fraction of 90\%,  we estimated that W0359 is not an equal-luminosity binary with a separation larger than $\sim$ 1.9 AU. For this object, the analysis is made more difficult by the small number of available stars in the field of view for generating a PSF. 

For W0535 our data is of sufficiently poor signal-to-noise that our simulations cannot achieve 90\% recovery over the separation limits studied. This is simply an outcome of the fact that the artificial star injection technique is not powerful when the signal-to-noise of the object detection  is $\lesssim$20 as is the case for W0535. Using the more relaxed 50\% recovery limit we estimate less stringent limits for the separation of a possible equal-luminosity binary companion of 1.9\,AU for W0535.

If these Y dwarfs are not binaries, the more plausible explanation for their over-luminosity is the presence of clouds. As discussed by \cite{TINNEY14}, the fact that some Y dwarfs show over-luminosity while others do not, could indicate different levels of cloud coverage between similar Y dwarfs or time-variable cloud coverage for the same object. Cloud  coverage has already been detected on brown dwarfs. \cite{FAHERTY14} have reported the presence of water ice clouds in the coolest brown dwarf known \citep[WISE J085510.83-071442.5;][]{LUHMAN14}, while photometric variability due heterogeneous cloud coverage has been reported in some T  \citep{BUENZLI12,APAI13,BUENZLI15} and Y \citep{CUSHING14A} dwarfs.

\subsection{Faint companions}

In general, both techniques deliver consistent results where they overlap at separations of $\sim$0.1\arcsec, with the photon counting technique extending to fainter potential companions at larger radial separations. Our best data on our brightest target (W0713) allow us to conclude that GSAOI--GeMS should able to resolve a brown dwarf companion with a contrast $<$ 4.4\,mag with respect to W0713 at separations beyond 1.66\,AU (0.18$\arcsec$) (Fig.~\ref{fig:limit1}). For W1541 and W1639 we rule out companions up to $\approx$3.5\,mag fainter than the known Y dwarf at separations beyond 0.5\,AU, while for W0359 and W0535 we rule out companions with a contrast $<$ 2.0 mag with respect to the Y dwarfs at separations greater than $\sim$2\,AU from the Y dwarf.

\subsection{Binary fraction and Mean Separation}

\begin{figure}
\centering
\hspace*{-0.1in}
\includegraphics[scale=0.5]{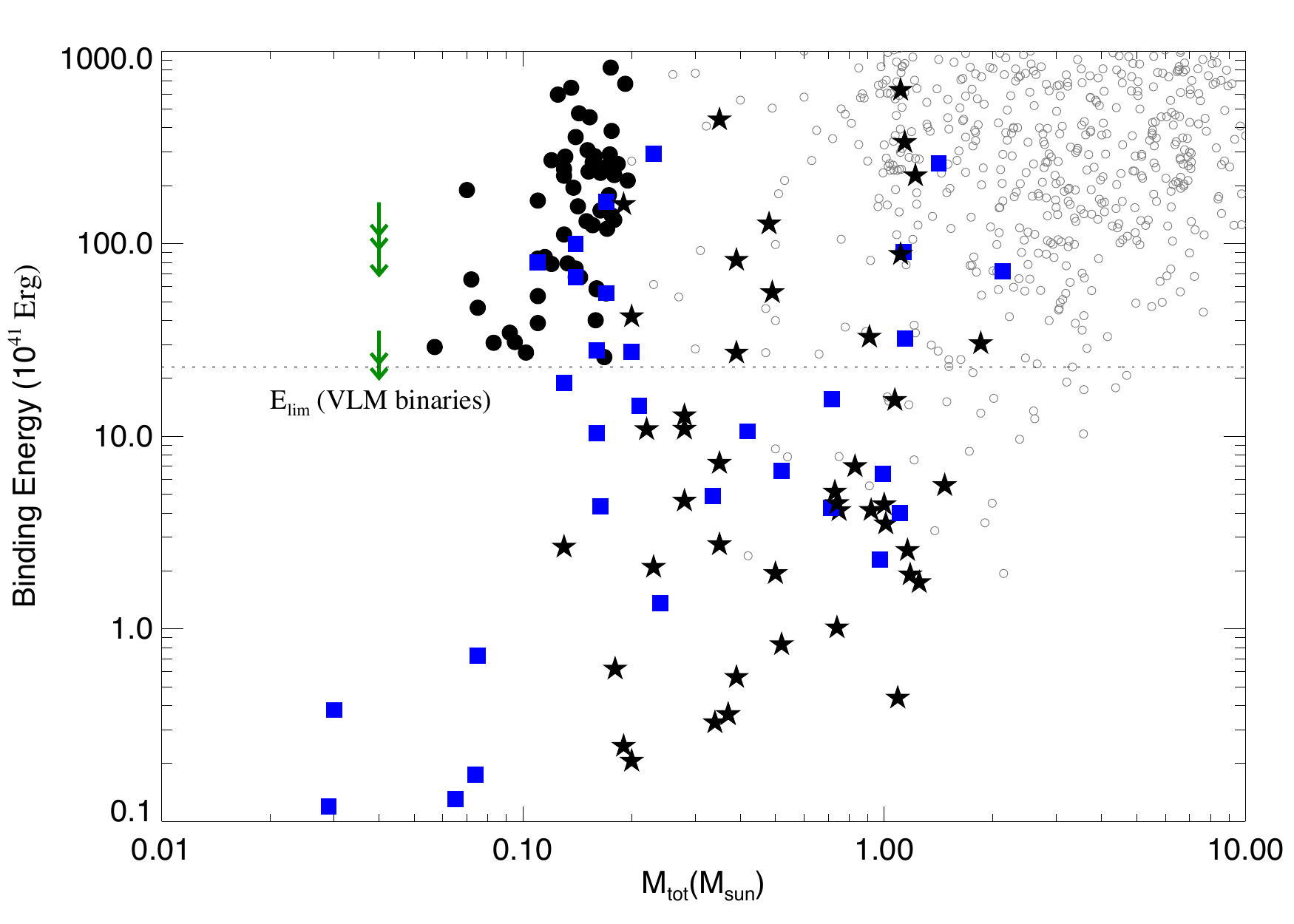}
\caption{Binding energy vs. total mass for known very low--mass systems. Objects marked with filled circles are tight very low–-mass systems (typically M$_{tot}$ $\sim$ 0.2 M$_\odot$ and $\rho$ $\sim$ 20 AU). Wide systems ( $\rho$ $\sim$ 100 AU) containing a ultra cool dwarf (UCD) companion are marked as five point stars. Those marked as squares are systems containing a tight or widely separated UCD with an age $<$ 500 Myr. Objects marked by open circles come from stellar companion catalogs. Our targets are marked with green arrows. The minimum binding energy corresponding to tight very low mass systems from \cite{CLOSE03, CLOSE07} and Burgasser et al. 2003 is labeled. We put the limits on each of the systems using an assumed mass of 20M$_{Jup}$ and an age of 5 Gyr \citep{DUPUY13}.}
\label{fig:BDenergy1}
\end{figure} 

\begin{figure}
\centering
\hspace*{-0.1in}
\includegraphics[scale=0.5]{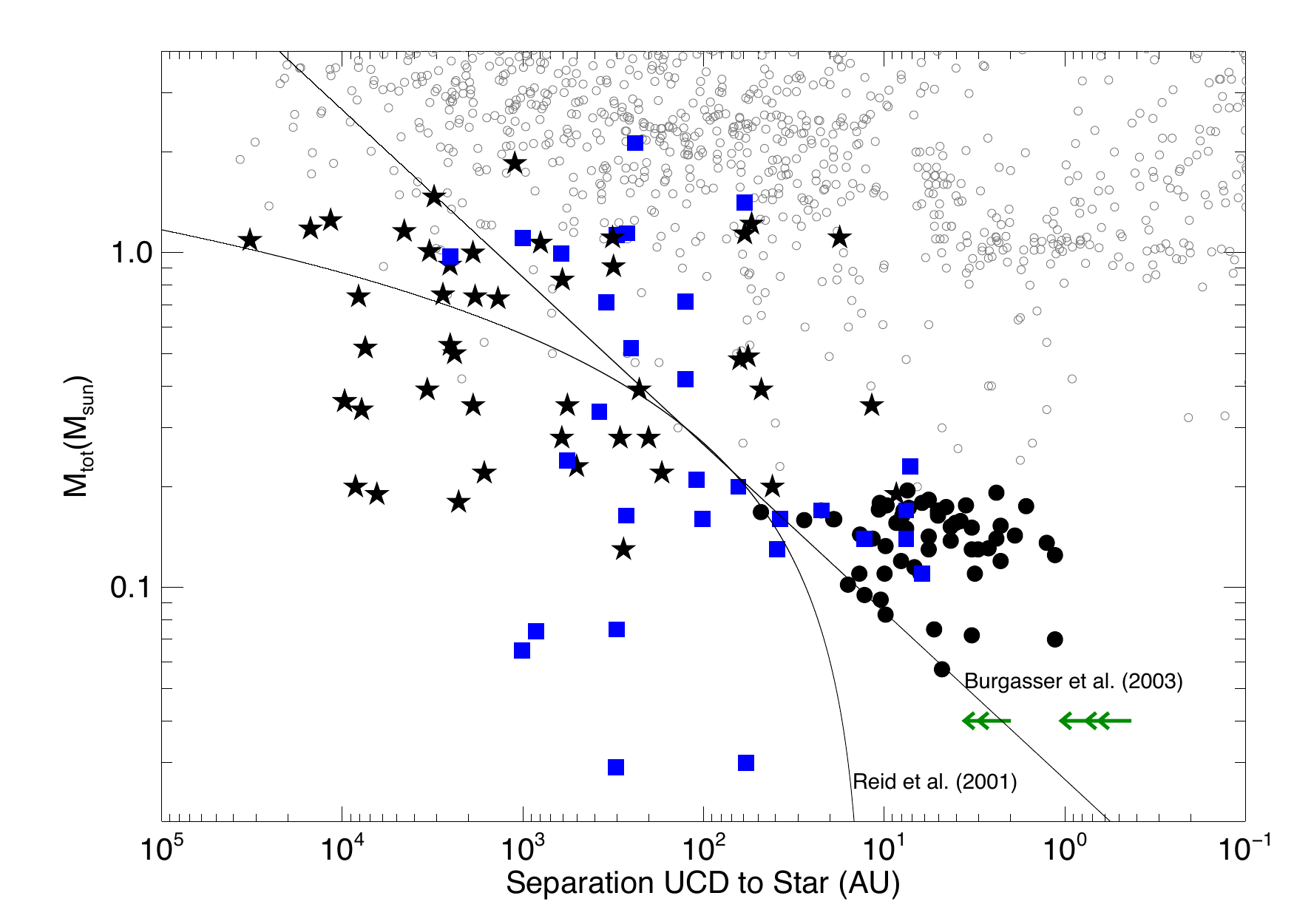}
\caption{Total mass vs separation. Symbols are described in Fig.~\ref{fig:BDenergy1}. The curves correspond to the empirical limits for the stability of binary systems set by \citep{REID01} and \citep{BURGASSER03}, where the stability area is located above and to the right side of each line. Both curves delineate the cutoff for the formation of stellar binary systems but \cite{BURGASSER03} is specific for a total mass of the system M$_{Tot} \leq$ 0.2$_\odot$.}
\label{fig:BDenergy2}
\end{figure} 

Observations show a decreasing binary fraction with later primary spectral type \citep[see e.g. ][and references therein.]{DUCHENE13}. A recent study of brown dwarf multiplicity \citep{ABERA14} infers an upper limit  of between 16\% and 25\%  for the binary fraction of brown dwarf companions to primaries of T5 and later. This is in agreement with previous brown dwarf multiplicity studies \citep{BURGASSER06,BURGASSER07}.

This trend in the binary fraction could be explained by either a mass dependence in the star formation process, or an observational bias. The latter is supported by the fact that multiplicity has been largely studied from resolved imaging programs, which are limited in resolution.  \cite{BURGASSER07} noted that the peak in the binary angular separation distribution was coincident with the resolution limit of Hubble Space Telescope and ground--based adaptive optics facilities, and suggested an undiscovered population of tight binaries (\textit{a} $\leq$ 1 AU).

There is scarce evidence for tight binary systems with mid--T to late--T dwarf primaries-- observed separations typically lie between $\sim$ 2 to 15 AU \citep{ABERA14}. However, recently \cite{DUPUY15} have extended multiplicity statistics into smaller separations by reporting a tight sub--stellar binary at the T/Y transition with a separation of $\sim$ 0.93 AU. 

In our sample of five Y dwarfs, we did not find evidence for companions down to separation limits of 0.3-1.9 AU, which is in agreement with the binary fraction estimates by \cite{ABERA14}. However, our sample is not large enough to confirm whether the decreasing binary fraction with later spectral type primaries is a real trend or an observational bias. 

In addition, we do not discount the possibility that some of the Y dwarfs of our sample harbor tighter binaries than our data can resolve, as the mean separation of known binaries also drops as a function of primary type. Very low--mass binaries are expected to be more bound as the gravitational potential well of each component drops as a function of the primary type \citep{CLOSE03, CLOSE07, BURGASSER03}.

In Fig.~\ref{fig:BDenergy1} we compare the estimated binding energy of our targets (for the assumption that they are unresolved binaries), with known binary systems collected in \cite{FAHERTY11} assuming near equal-mass companions of 20 M$_{Jup}$ \citep[from][]{DUPUY13} for our targets. The objects of our sample fall within the binding energy limitation set by known tight low-mass (M$_{tot}$ $<$ 0.2 M$_\odot$) multiples (the dotted line in the figure).

Fig.~\ref{fig:BDenergy2} shows the M$_{tot}$ versus separation for the same companion systems described in Fig.~\ref{fig:BDenergy1}, along with the maximum separations allowed to keep the binary system stable, as suggested by both \cite{REID01} and \cite{BURGASSER03}. In both cases, these empirical limits were derived from the binary systems known at that time. As has been noted by \cite{DITHAL10} and \cite{FAHERTY10}, these cutoffs break down for more massive and widely separated systems as well as very young ($<$ 10 Myr) systems. Nonetheless, the trends predicted by \citep{REID01} and \citep{BURGASSER03} provide a means to extrapolate the properties observed for more massive binary systems, to the lower masses relevant for our Y dwarf observations. This suggests that the upper-limits we observe for binary separation in our five Y dwarfs are not inconsistent (based on binding energy and stability arguments) with the binary separations seen at larger masses.

Finally, although no companions to these Y dwarfs were discovered, we have reached limiting angular separations as small as 0.04$\arcsec$. GSAOI--GeMS therefore, is an excellent instrument to expand multiplicity statistics of the coldest brown dwarfs.

\section{Summary and Conclusions}

We have observed five WISE brown dwarfs with the Gemini GeMS Multi-Conjugate Adaptive Optics System to identify ultra-cool companions and have implemented two methods to compute sensitivities as a function of separation and luminosity. Combining the results computed by two different techniques we conclude:

\begin{itemize}
\item{We detect no binary companions to the five Y dwarfs observed.}
\item{None of these Y dwarfs are equal-mass/equal-luminosity binaries with separations larger than $\sim$ 0.5-1.9 AU. Our best data are for W1541, where artificial star injection  (at a recovery fraction of 90\%), shows no evidence of an equal-mass/equal-luminosity binary at separations down to 0.5 AU (0.08$\arcsec$).}
\item{GSAOI-GeMS would be able to detect binary companions as much as $\sim$ 4.4 mag fainter than the known Y dwarf at separations beyond 0.08$\arcsec$.}
\end{itemize}

Although no  binary companions to Y dwarfs were detected, these data probe an interesting range of orbital separations for these nearby Y dwarfs and demonstrated the power of GSAOI-GeMS for this science.\\

\acknowledgments
We gratefully acknowledge the support of ARC Australian Professorial Fellowship grant DP0774000 and ARC Discovery Outstanding Researcher Award DP130102695. D. Opitz is also supported by CONICYT Becas Chile 72130434. This paper is based on observations obtained at the Gemini Observatory, which is operated by the Association of Universities for Research in Astronomy, Inc., under a cooperative agreement with the NSF on behalf of the Gemini partnership: the National Science Foundation (United States), the National Research Council (Canada), CONICYT (Chile), the Australian Research Council (Australia), Minist\'{e}rio da Ci\^{e}ncia, Tecnologia e Inova\c{c}\~	{a}o (Brazil) and Ministerio de Ciencia, Tecnolog\'{i}a e Innovaci\'{o}n Productiva (Argentina). Time has been awarded through Australia and USA via programs GS-2013B--Q--26, GS--2014A--Q--4, GS--2014B--Q--32 and also via Guaranteed Time programs GS--2014A--C--3 and GS--2014B--C--1. We would like to acknowledge the high standard of support offered by the Gemini queue observing team who acquired most of the data used in this paper. The authors would like to especially acknowledge the extraordinary quality of the instrument delivered for use by our team (and others) by the GSAOI Principal Investigator Professor Peter McGregor and his team at the Australian National University. We thank Dr R. Sharp for his assistance in acquiring data for this program during GSAOI Guaranteed Time. We also thank Dr. D. Wright for helpful comments and suggestions on this manuscript. 

{\it Facilities:} \facility{GEMINI (GeMS--GSAOI)}.\\


\begin{thebibliography}{}
\bibitem[Aberasturi et al.(2014)]{ABERA14} Aberasturi, M., Burgasser, A.~J., Mora, A., et al.\ 2014, \aj, 148, 129 
\bibitem[Apai et al.(2013)]{APAI13} Apai, D., Radigan, J., Buenzli, E., et al.\ 2013, \apj, 768, 121 
\bibitem[Bouy et al.(2006)]{BOUY06} Bouy, H., Moraux, E., Bouvier, J., et al.\ 2006, \apj, 637, 1056

\bibitem[Buenzli et al.(2012)]{BUENZLI12} Buenzli, E., Apai, D., Morley, C.~V., et al.\ 2012, \apjl, 760, LL31 
\bibitem[Buenzli et al.(2015)]{BUENZLI15} Buenzli, E., Saumon, D., Marley, M.~S., et al.\ 2015, \apj, 798, 127 
\bibitem[Burgasser et al.(2003)]{BURGASSER03} Burgasser, A.~J., Kirkpatrick, J.~D., Reid, I.~N., et al.\ 2003, \apj, 586, 512
\bibitem[Burgasser et al.(2006)]{BURGASSER06} Burgasser, A.~J., Kirkpatrick, J.~D., Cruz, K.~L., et al.\ 2006, \apjs, 166, 585
\bibitem[Burgasser et al.(2007)]{BURGASSER07} Burgasser, A.~J., Reid, I.~N., Siegler, N., et al.\ 2007, Protostars and Planets V, 427 
\bibitem[Carrasco et al.(2012)]{CARRASCO12} Carrasco, E.~R., Edwards, M.~L., McGregor, P.~J., et al.\ 2012, \procspie, 8447, 84470N 

\bibitem[Close et al.(2003)]{CLOSE03} Close, L.~M., Siegler, N., Freed, M., \& Biller, B.\ 2003, \apj, 587, 407 
\bibitem[Close et al.(2007)]{CLOSE07} Close, L.~M., Zuckerman, B., Song, I., et al.\ 2007, \apj, 660, 1492 
\bibitem[Cushing et al.(2011)]{CUSHING11} Cushing, M.~C., Kirkpatrick, J.~D., Gelino, C.~R., et al.\ 2011, ApJ, 743, 50
\bibitem[Cushing et al.(2014)]{CUSHING14} Cushing, M.~C., Kirkpatrick, J.~D., Gelino, C.~R., et al.\ 2014, \aj, 147, 113
\bibitem[Cushing et al.(2014a)]{CUSHING14A} Cushing, M., Hardegree-Ullman, K., \& Trucks, J.\ 2014, American Astronomical Society Meeting Abstracts 223, 425.08 
\bibitem[Delfosse et al.(2004)]{DELFOSSE04} Delfosse, X., Beuzit, J.-L., Marchal, L., et al.\ 2004, Spectroscopically and Spatially Resolving the Components of the Close Binary Stars, 318, 166
\bibitem[d'Orgeville et al.(2012)]{ORGEVILLE12} d'Orgeville, C., Diggs, S., Fesquet, V., et al.\ 2012, \procspie, 8447, 84471Q
\bibitem[Dhital et al.(2010)]{DITHAL10} Dhital, S., West, A.~A., Stassun, K.~G., \& Bochanski, J.~J.\ 2010, \aj, 139, 2566 
\bibitem[Duch{\^e}ne \& Kraus(2013)]{DUCHENE13} Duch{\^e}ne, G., \& Kraus, A.\ 2013, \araa, 51, 269 
\bibitem[Dupuy et al.(2009)]{DUPUY09} Dupuy, T.~J., Liu, M.~C., \& Ireland, M.~J.\ 2009, \apj, 692, 729
\bibitem[Dupuy \& Kraus(2013)]{DUPUY13} Dupuy, T.~J., \& Kraus, A.~L.\ 2013, Science, 341, 1492  
\bibitem[Dupuy et al.(2015)]{DUPUY15} Dupuy, T.~J., Liu, M.~C., \& Leggett, S.~K.\ 2015, \apj, 803, 102 
\bibitem[Duquennoy \& Mayor(1991)]{DUQUENNOY91} Duquennoy, A., \& Mayor, M.\ 1991, \aap, 248, 485 
\bibitem[Gelino et al.(2011)]{GELINO2011} Gelino, C.~R., Kirkpatrick, J.~D., Cushing, M.~C., et al.\ 2011, \aj, 142, 57
\bibitem[Faherty et al.(2010)]{FAHERTY10} Faherty, J.~K., Burgasser, A.~J., West, A.~A., et al.\ 2010, \aj, 139, 176 
\bibitem[Faherty et al.(2011)]{FAHERTY11} Faherty, J.~K., Burgasser, A.~J., Bochanski, J.~J., et al.\ 2011, \aj, 141, 71 
\bibitem[Faherty et al.(2014)]{FAHERTY14} Faherty, J.~K., Tinney, C.~G., Skemer, A., \& Monson, A.~J.\ 2014, \apjl, 793, LL16 
\bibitem[Kirkpatrick et al.(2011)]{KIRK11} Kirkpatrick, J.~D., Cushing, M.~C., Gelino, C.~R., et al.\ 2011, \apjs, 197,19 
\bibitem[Kirkpatrick et al.(2012)]{KIRK12} Kirkpatrick, J.~D., Gelino, C.~R., Cushing, M~C., et al.\ 2012, ApJ, 753, 156
\bibitem[Kirkpatrick et al.(2013)]{KIRK13} Kirkpatrick, J.~D., Cushing, M.~C., Gelino, C.~R., et al.\ 2013, \apj, 776, 128 
\bibitem[Koerner et al.(1999)]{KOERNER99} Koerner, D.~W., Kirkpatrick, J.~D., McElwain, M.~W., \& Bonaventura, N.~R.\ 1999, \apjl, 526, L25
\bibitem[Konopacky(2013)]{KONOPACKY13} Konopacky, Q.~M.\ 2013, \memsai, 84, 1005 
\bibitem[Konopacky et al.(2010)]{KONOPACKY10} Konopacky, Q.~M., Ghez, A.~M., Barman, T.~S., et al.\ 2010, \apj, 711, 1087 
\bibitem[Leggett et al.(2015)]{LEGGETT15} Leggett, S.~K., Morley, C.~V., Marley, M.~S., \& Saumon, D.\ 2015,\apj, 799, 37 
\bibitem[Liu et al.(2011)]{LIU11} Liu, M.~C., Delorme, P., Dupuy, T.~J., et al.\ 2011, \apj, 740, 108
\bibitem[Liu et al.(2012)]{LIU12} Liu, M.~C., Dupuy, T.~J., Bowler, B.~P., Leggett, S.~K., \& Best, W.~M.~J.\ 2012, \apj, 758, 57 
\bibitem[Luhman et al.(2011)]{LUHMAN11} Luhman, K.~L., Burgasser, A.~J., \& Bochanski, J.~J.\ 2011, \apjl, 730, L9
\bibitem[Luhman(2014)]{LUHMAN14} Luhman, K.~L.\ 2014, \apj, 781, 4 
\bibitem[Mart{\'{\i}}n et al.(1998)]{MARTIN98} Mart{\'{\i}}n, E.~L., Basri, G., Brandner, W., et al.\ 1998, \apjl, 509, L113 
\bibitem[McGregor et al.(2004)]{MCGREGOR04} McGregor, P., Hart, J., Stevanovic, D., et al.\ 2004, \procspie, 5492, 1033 
\bibitem[Morley et al.(2012)]{MORLEY12} Morley, C.~V., Fortney, J.~J., Marley, M.~S., et al.\ 2012, \apj, 756, 172
\bibitem[Neuh{\"a}user et al.(2002)]{NEUHAUSER02} Neuh{\"a}user, R., Brandner, W., Alves, J., Joergens, V., \& Comer{\'o}n, F.\ 2002, \aap, 384, 999 
\bibitem[Pinfield et al.(2014)]{PINFIELD14} Pinfield, D.~J., Gromadzki, M., Leggett, S.~K., et al.\ 2014, \mnras, 444, 1931
\bibitem[Reid \& Gizis(1997)]{REID97} Reid, I.~N., \& Gizis, J.~E.\ 1997, \aj, 113, 2246
\bibitem[Reid et al.(2001)]{REID01} Reid, I.~N., Gizis, J.~E., 
Kirkpatrick, J.~D., \& Koerner, D.~W.\ 2001, \aj, 121, 489 
\bibitem[Schneider et al.(2015)]{SCHNEIDER15} Schneider, A.~C., Cushing, M.~C., Kirkpatrick, J.~D., et al.\ 2015, \apj, 804, 92
\bibitem[Reid et al.(2008)]{REID08} Reid, I.~N., Cruz, K.~L., Burgasser, A.~J., \& Liu, M.~C.\ 2008, \aj, 135, 580 
\bibitem[Stetson(1987)]{STETSON87} Stetson, P.~B.\ 1987, \pasp, 99, 191 
\bibitem[Tinney et al.(2012)]{TINNEY12} Tinney, C.~G., Faherty, J.~K., Kirkpatrick, J.~D., et al.\ 2012, \apj, 759, 60 
\bibitem[Tinney et al.(2014)]{TINNEY14} Tinney, C.~G., Faherty, J.~K., Kirkpatrick, J.~D., et al.\ 2014, \apj, 796, 39 
\bibitem[Todorov et al.(2014)]{TODOROV14} Todorov, K.~O., Luhman, K.~L., Konopacky, Q.~M., et al.\ 2014, \apj, 788, 40
\bibitem[Wright et al.(2010)]{WRIGHT10} Wright, E.~L., Eisenhardt, P.~R.~M., Mainzer, A.~K., et al.\ 2010, \aj, 140, 1868 
\end{thebibliography}
\end{document}